\documentclass[prb,aps,twocolumn,showpacs]{revtex4}

\usepackage{epsfig}

\begin{document}

\title{Local spin fluctuations in iron-based superconductors: $^{77}$Se and $^{87}$Rb NMR measurements of Tl$_{0.47}$Rb$_{0.34}$Fe$_{1.63}$Se$_2$}

\author{Long Ma}
\author{G. F. Ji}
\author{Jia Dai}
\author{J. B. He}
\author{D. M. Wang}
\author{G. F. Chen}
\author{Bruce Normand}
\author{Weiqiang Yu}
\email{wqyu_phy@ruc.edu.cn}
\affiliation{Department of Physics, Renmin University of China, Beijing 
100872, China}

\date{\today}

\pacs{74.70.-b, 76.60.-k}

\begin{abstract}

We report nuclear magnetic resonance (NMR) studies of the intercalated 
iron selenide superconductor (Tl, Rb)$_{y}$Fe$_{2-x}$Se$_2$ ($T_c = 32$ K). 
Single-crystal measurements up to 480 K on both $^{77}$Se and $^{87}$Rb 
nuclei show a superconducting phase with no magnetic order. The Knight 
shifts $K$ and relaxation rates $1/T_1T$ increase very strongly with 
temperature above $T_c$, before flattening at 400 K. The quadratic 
$T$-dependence and perfect proportionality of both $K$ and $1/T_1T$ data 
demonstrate their origin in paramagnetic moments. A minimal model for 
this pseudogap-like response is not a missing density of states but two 
additive contributions from the itinerant electronic and local magnetic 
components, a framework unifying the $K$ and $1/T_1 T$ data in many 
iron-based superconductors. 

\end{abstract}

\maketitle

The phenomenon of the pseudogap, an energy scale marking the 
loss of spin-fluctuation contributions in many physical 
processes,\cite{Warren_PG, Alloul_PG} is ubiquitous in underdoped 
cuprate superconductors. The origin of the pseudogap and its 
connection to high-temperature superconductivity have remained 
controversial questions for over two decades. In the iron pnictide 
superconductors,\cite{Hosono_Jacs_130_3296, Chen_Nature_453_761, 
Ren_CPL_12_105, Chen_PRL_100_247002} the susceptibility and Knight 
shift in the normal state have been found to increase with 
temperature,\cite{ChenXHSUS, CanfieldSUS, BuechnerSUS, Ning_PRL_104} and 
while a pseudogap has been proposed\cite{Nakai_JPSJ, Ning_PRL_104} 
in comparison with cuprates, the evidence is far from conclusive.

The recently discovered intercalated iron selenide superconductors, 
A$_{y}$Fe$_{2-x}$Se$_2$, have transition temperatures $T_c > 30$ 
K.\cite{Guo_PRB_82_180520, Mizuguchi_10124950, Chen_CM_10125525, FangMH_CM_10125236, 
Chen_CM_11010789} Structural experiments indicate a 
distinctive $\sqrt{5} \times \! \sqrt{5}$ ordering pattern of Fe vacancies
in the Fe$_4$Se$_5$ layers.\cite{BaoW_11014882, Ye_11022882} Accompanying 
this structure is an equally distinctive block-spin antiferromagnetic (AFM) 
order with a very high N\'eel temperature.\cite{Bao_11020830, Ye_11022882, 
ChenXH_11022783, Shermadini_11011873} By contrast, NMR reveals a 
superconducting state with only weak low-energy spin fluctuations and 
no evidence for magnetic order,\cite{YuW_11011017, ImaiT_11014967, Ma11023888} 
which is consistent with angle-resolved photoemission spectroscopy
\cite{Feng_CM_10125980, ZhouXJ_11014556, Ding_10126017}. 
The substantial increases of $K$ and $1/T_1T$ reported above $T_c$ in 
Refs.~[\onlinecite{YuW_11011017, ImaiT_11014967, Ma11023888}]
suggest a strong pseudogap-like effect, and this would set strict constraints on 
any theoretical models. 

In this Letter, we present an extensive $^{77}$Se and $^{87}$Rb NMR 
investigation of superconducting (Tl,Rb)$_{y}$Fe$_{2-x}$Se$_2$ single 
crystals at temperatures up to 480 K. The Knight shifts and relaxation 
rates on both the $^{77}$Se and $^{87}$Rb sites rise by factors of five to 
20 above $T_c$, before levelling off towards 400 K. 
At low temperatures, our measured response is due to itinerant electrons
forming a Fermi liquid. Above $T_c$, we have previously identified a 
quadratic temperature-dependence of the Knight shift\cite{Ma11023888}.
Here we further demonstrate that the rise of both the Knight shift and
the $1/T_1T$ follows very accurately a $T^2$ 
dependence, indicative of fluctuating paramagnetic moments. The leveling 
off above 400 K suggests that the energy scale of the spin fluctuations is around 35 meV.
Thus our data, whose form is reminiscent of a pseudogap behavior, is best 
explained by a two-component model. This situation is quite different from cuprates and provides 
a general basis for interpreting $K$ and $1/T_1 T$ data in all the Fe 
superconductors. 

The (Tl,Rb)$_{y}$Fe$_{2-x}$Se$_2$ single crystals were synthesized by the 
Bridgeman method.\cite{Chen_CM_11010789} Inductively coupled plasma
atomic emission spectroscopy measurements give a nominal chemical 
composition of Tl$_{0.47}$Rb$_{0.34}$Fe$_{1.63}$Se$_2$. Several crystals, 
all of dimensions 5$\times$4$\times$1 mm$^3$ and $T_c =$ 32$\pm$1 K,
were used in our NMR experiments, and all gave consistent results. A 
specially adapted NMR probe of our own construction was used to access 
both low and high temperatures. We have performed $^{87}$Rb ($^{87}\gamma_n
 = 13.931$ MHz/T) and $^{77}$Se ($^{77}\gamma_n = 8.131$ MHz/T) 
measurements\cite{Me2Se} in a magnetic field of 11.62 T, with the sample 
placed on a rotator to change the field orientation. The spectral linewidth 
is approximately 25 kHz for $^{87}$Rb and 20 kHz for $^{77}$Se at $T =$ 40 K. 
The spin-lattice relaxation rate $1/T_1$ is measured by the inversion-recovery 
method, and a single $T_1$ component is obtained. 

\begin{figure}
\includegraphics[width=8cm]{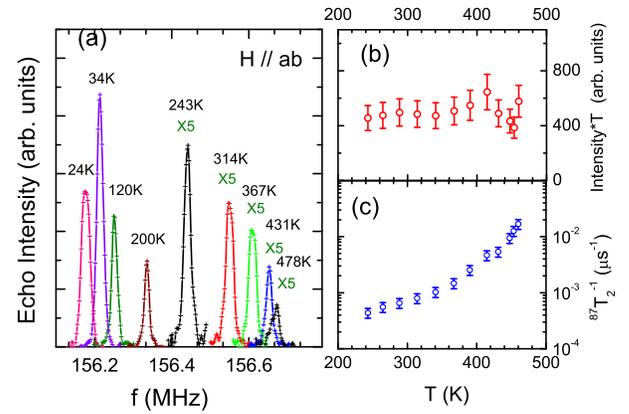}
\caption{\label{spec2}(color online) (a) $^{87}$Rb NMR spectra in a field 
of 11.62 T applied in the crystalline ($ab$) plane, for $24 < T < 478$ K. 
(b) Integrated spectral weight as a function of temperature. (c) $^{87}$Rb 
spin-spin relaxation rate, $1/^{87}T_2$, up to 460 K.}  
\end{figure}

Figure \ref{spec2}(a) shows the NMR spectra of the $^{87}$Rb central transition 
at different temperatures. The central frequency shifts upward gradually with 
temperature. The spin-spin relaxation rate $1/^{87}T_2$ also increases with 
temperature [Fig.~\ref{spec2}(c)], until the NMR spectrum becomes undetectable 
beyond 480 K, where $^{87}T_2$ becomes too short. After correction for $T_2$ 
effects, the spectral weight is conserved over the full temperature range 
of our studies [Fig.~\ref{spec2}(b)]. For $^{77}$Se, $T_2$ also becomes 
shorter with increasing temperature (data not shown), and the $^{77}$Se 
spectrum is not detectable above 420 K.  

\begin{figure}
\includegraphics[width=8.5cm]{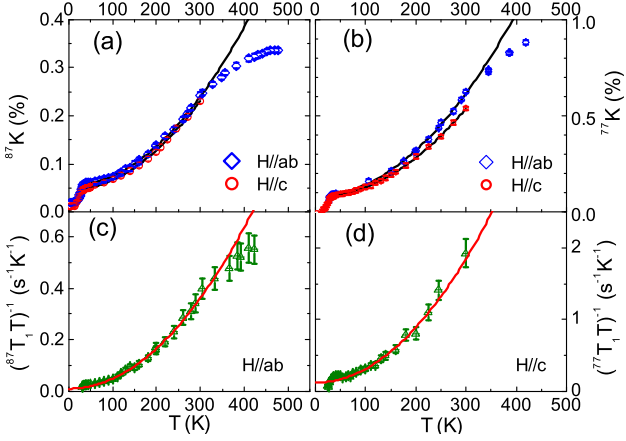}
\caption{\label{kslrr}(color online) Temperature-dependences of (a)
$^{87}K$, (b) $^{77}K$, (c) $1/^{87}T_1T$, and (d) $1/^{77}T_1T$ in a
magnetic field of 11.62 T. The solid lines are fits to the functions 
$K(T) = a + b T^2$ and $1/T_1T = c + d T^2$ from $T_c$ to 300 K.}
\end{figure}

In Figs.~\ref{kslrr}(a) and (b), we show $^{87}K$ and $^{77}K$ in fields 
applied along the crystalline $c$ axis and in the ($ab$) plane, and over 
a very wide range of temperature. The Knight shifts for the two field 
orientations are nearly identical. The two nuclei also have rather similar 
temperature-dependences. Focusing first on the region below $T_c$, 
in general the Knight shift $K = K_c + K_s$ has both chemical ($K_c$) and 
spin ($K_s$) contributions. $K_c$ is independent of temperature while $K_s$ 
approaches zero for a singlet superconductor. Singlet superconductivity is 
therefore demonstrated quite unambiguously by the sharp drop below 
$T_c$.\cite{YuW_11011017} For temperatures far below $T_c$, $K(T) 
\approx 0$ suggests that $K_c$ is very small in the A$_{y}$Fe$_{2-x}$Se$_2$ 
materials,\cite{Ma11023888} and this contribution will be neglected hereafter.

Turning to the normal-state behavior of $K(T)$, it is almost isotropic 
and comparable on both $^{77}$Se and $^{87}$Rb nuclei. It increases strongly 
from $T_c$ to 400 K [Figs.~2(a) and (b)] before flattening out. This form 
is reminiscent of a pseudogap, where spin fluctuations contributing to 
$K(T)$ are suppressed below a characteristic energy scale, $\Delta_{\rm pg}$. 
In the simplest pseudogap scenario, $K(T)$ at low $T$ should contain a 
thermally activated contribution and $\Delta_{\rm pg}$ can be extracted from 
the data.\cite{Bankay_PRB_50_6416, Walstedybook}

\begin{figure}
\includegraphics[width=6.5cm]{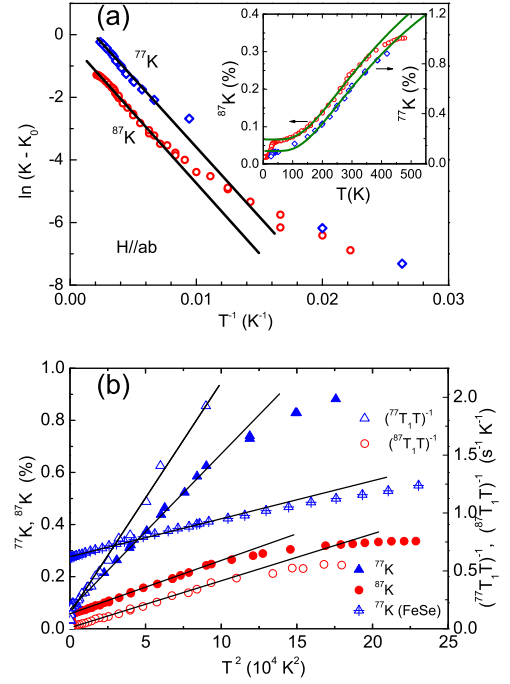}
\caption{\label{kn3}(color online)  (a) $T$-dependence for $^{87}K$ (circles) 
and $^{77}K$ (diamonds), shown as $\ln (K(T) - K_0)$ against $1/T$ (see text). 
Inset: Knight shifts compared on linear axes. Solid lines are fits to the 
activated form. (b) Knight shifts and relaxation rates for both nuclei, 
shown as functions of $T^2$. The $^{77}K$ data of FeSe are adapted 
from Ref.~[\onlinecite{Imai_prl_102_177005}]. }
\end{figure}

We test for such activated behavior in Fig.~\ref{kn3}(a). The expressions 
$^{87}K(T) = 0.066\% + 0.946\% e^{-510 {\rm K}/T}$ and $^{77}K(T) = 0.1\% + 
2.73\% e^{-502 {\rm K}/T}$ provide adequate fits over the full range from 
$T_c$ to 350 K [inset, Fig.~\ref{kn3}(a)], and suggest an effective gap 
scale around 500 K. However, $K(T)$ levels off around 400 K, and a 
detailed inspection of the low-$T$ regime in a standard activation plot 
[Fig.~3(a)] shows clearly that $K(T)$ does not follow this form below 
150 K. The fitting contains no evidence for a pseudogap energy scale and 
the data require an alternative explanation.

In fact an excellent fit to all of the Knight-shift data below 300 K 
[Figs.~2(a) and (b)], is provided by the function $K(T) = a + b T^2$, 
where $^{77}K(T) = 0.079\% + 5.978 \! \times \! 10^{-6}\% T^2/K^2$ and
$^{87}K(T) = 0.055\% + 2.706 \! \times \! 10^{-6}\% T^2/K^2$. 
For comparison with the activated scenario, the data and fit are 
reproduced in Fig.~\ref{kn3}(b). Only above 300 K do the measured 
Knight shifts deviate from the $T^2$ behavior as the data begin to 
saturate. Even more remarkably, the relaxation rates for both nuclei, shown 
in Figs.~\ref{kslrr}(c) and (d) and also analyzed in Fig.~\ref{kn3}(b), have 
precisely the same behavior from $T_c$ to 300 K, with
$1/^{77}T_1T = 0.136s^{-1}K^{-1} + 2.329 \! \times \! 10^{-5}T^2s^{-1}K^{-3}$ and
$1/^{87}T_1T = 0.014s^{-1}K^{-1} + 3.905 \! \times \! 10^{-6}T^2s^{-1}K^{-3}$. 
The quadratic $T$-dependence of both quantities is completely unambiguous, 
and we discuss its physical meaning below. 

\begin{figure}
\includegraphics[width=8cm]{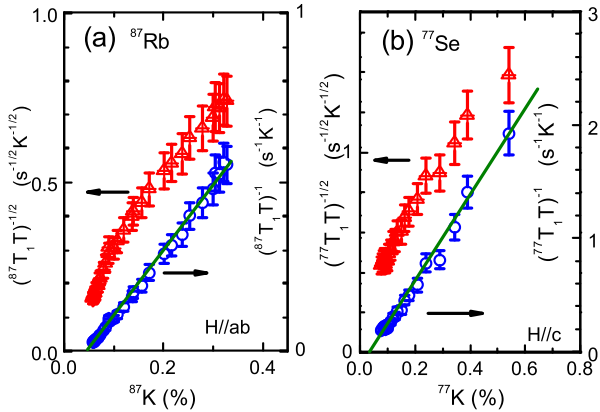}
\caption{\label{t1ks}(color online) $(T_1T)^{-\frac12}$ (left axis) 
and $(T_1T)^{-1}$ (right) against $K$ for (a) $^{87}$Rb with $T_c 
\le T \le 425$ K and (b) $^{77}$Se with $T_c \le T \le 300$ K. 
Straight lines are guides to the eye.}  
\end{figure}

We further analyze our data by comparing $K(T)$ to $(T_1T)^{-1}$ with 
temperature as the implicit parameter. A system with Fermi-liquid behavior 
follows the Korringa relation, $(T_1T)^{-1/2} \propto K(T)$.\cite{slichterbook} 
A linear relation between $K(T)$, which measures the $q = 0$ response, and 
$(T_1T)^{-1}$, which is the $q$-integrated response, is the hallmark of 
relaxation processes dominated by local spin fluctuations. It is clear from 
Figs.~\ref{t1ks}(a) and (b) that the Korringa relation is not followed over 
any of the temperature range, whereas good linear proportionality is obtained 
between $K(T)$ and $(T_1T)^{-1}$. We conclude that spin fluctuations are the 
predominant relaxation mechanism above $T_c$, and that these are very local 
in nature as there is no significant contribution from particular wave 
vectors ${\bf Q} \ne 0$. 

We stress here that the fitting lines in Figs.~\ref{t1ks}(a) and 
(b) do not pass through the origin. Both $K(T)$ and $1/T_1T$, for both 
nuclei, approach constant values as $T \rightarrow T_c$. Such constant 
terms above $T_c$ usually indicate an additional Fermi-liquid contribution 
from itinerant electrons,\cite{Abragambook} which is lost as the system 
becomes superconducting below $T_c$ (Fig.~2). The Fermi-liquid contribution 
may be extracted by writing $K(T) = K_0 + f(T)$ and $1/T_1T = (1/T_1T)_0 + 
f'(T)$, where the $T$-dependence, contained only in $f(T)$ and $f'(T)$, 
is quadratic. This explicit separation of itinerant-electron and 
spin-fluctuation contributions is completed by identifying the two 
ratios $(1/T_1T)_0/K_0$ and $f'(T)/f(T)$, whose respective values are 
34 s$^{-1}$K$^{-1}$  and 170 s$^{-1}$K$^{-1}$ for $^{87}$Rb, and 190 
s$^{-1}$K$^{-1}$ and 350 s$^{-1}$K$^{-1}$ for $^{77}$Se.

We conclude the analysis of our data with two further comments on the 
Knight shift. First, the anisotropy is very small, with $^{77}K_{ab}/ 
^{77}K_c \approx$ 1.1 and $^{87}K_{ab}/^{87}K_c \approx$ 1.0 at 300 K 
(Fig.~\ref{kslrr}). Second, the ratio $^{87}K/^{77}K \approx 0.4$ at 
300 K indicates that the amplitudes of the hyperfine fields, which are 
produced by the Fe moments, are quite similar at the in-plane Se site 
and the interlayer Rb site. These results suggest a strong interlayer 
magnetic coupling in the intercalated iron selenides. In the iron 
pnictides, the hyperfine fields are usually very much smaller at the 
interlayer nuclei than at the $^{75}$As site.\cite{MaNaFeAs, Imai_08044026}

Before discussing the two-component interpretation of our NMR measurements, 
we discuss the necessity for a two-phase scenario in the A$_{y}$Fe$_{2-x}$Se$_2$ 
materials. The block-AFM state of the $\sqrt{5} \times \! \sqrt{5}$ structure 
has a high $T_N \approx$ 550 K and a large magnetic moment of order 
3$\mu_{\rm B}$ per Fe site. These neutron-diffraction results\cite{Bao_11020830, 
Ye_11022882} have been verified by bulk magnetization,\cite{ChenXH_11022783} 
muon spin resonance ($\mu$SR),\cite{Shermadini_11011873} 
M\"ossbauer,\cite{Ryan_11030059} two-magnon 
Raman,\cite{Zhang_11062706} and inelastic neutron scattering 
measurements \cite{Wang_11054675}. However, the superconducting phase 
observed by NMR is clearly paramagnetic, because the large line-splitting 
of the $^{77}$Se spectra, expected for a strongly AFM state, is completely 
absent. Further, the Knight shift would not be nearly isotropic in the 
presence of AFM order. 

We are forced to conclude that the electronic matter in A$_{y}$Fe$_{2-x}$Se$_2$ 
is a two-phase system. The AFM phase is thought from $\mu$SR and neutron 
diffraction experiments to constitute at least 90\% of this system. The 
paramagnetic, metallic phase we observe, which makes up the remainder, may 
be a consequence of regions within the single crystals with different Fe 
vacancy content or vacancy disorder. The paramagnetic phase is manifestly 
superconducting. Whether the AFM phase is an insulator,\cite{Zhang_11024575} 
while the paramagnetic phase percolates throughout the system to give the 
observed bulk superconducting properties, or is a metal hosting a microscopic 
coexistence of magnetism and superconductivity, remains an open question that 
NMR cannot address. 

Returning now to the NMR response of the paramagnetic phase, the minimal 
model describing our experimental data is simply one with two additive 
components. One is the metallic behavior of the itinerant electrons, which 
become superconducting below $T_c$ and appear above $T_c$ as a $T$-independent 
constant. The origin of the quadratic $T$-dependence above $T_c$ lies in local 
fluctuations of the paramagnetic Fe moments, which may be as large as the 
3$\mu_{\rm B}$ observed in the AFM phase. These moments have only short-range 
order, with local AFM fluctuations, and while the system is close to the 
long-range order of the neighboring magnetic phase, this is prevented by 
some combination of non-stoichiometry, vacancy disorder, and the effects 
of the itinerant electrons. 

In studies of quantum AFMs, paramagnetic spins in a critical regime with 
no spin gap give a linear contribution to the susceptibility in two 
dimensions (2D) and a quadratic $T$-dependence in 3D.\cite{Normand_PRB} 
Similar considerations have been applied in pnictide superconductors to 
discuss the linear susceptibility in a 2D model\cite{Zhang_EPL} for 
1111 and 122 systems. The large values of $^{87}K(T)$ we measure are a 
definite indication of strong interlayer magnetic coupling and 3D nature 
in the A$_y$Fe$_{2-x}$Se$_2$ materials, and hence are fully consistent with 
a quadratic spin-fluctuation contribution. The temperature at which our 
data begin to turn over, $T \sim 400$ K, also agrees well with the 
predominant energy scale for local spin-exchange processes, $|SJ_1| = 
36$ meV, deduced in Ref.~[\onlinecite{Wang_11054675}]; this further 
strengthens the evidence in favor of fluctuating local moments rather 
than electronic processes.

While it may be possible to conceive of a pseudogap model with an exact 
constant contribution, exactly quadratic $T$-dependence, exact proportionality 
of $K$ and $1/T_1T$ over the entire temperature range above $T_c$, and the 
observed dramatic increases in these quantities, this would require very careful 
fine-tuning. Thus we find a two-component scenario to be much more likely, 
and it is certainly the minimal model for our results. Its foundation 
lies in the fact that Fe has active conduction and valence bands, the former 
providing itinerant electrons and the latter forming effective local 
moments.\cite{Wu_PRL, Medici_PRL_102} A two-component model has also been 
invoked to explain EPR and NMR data in FeSe$_x$Te$_{1-x}$\cite{Arcon_PRB_82} 
and NMR data in 111 materials.\cite{Jeglic_LiFeAs_NMR, Li_JPSJ_79} This 
situation is not at all similar to the superconducting cuprates, where 
the NMR response is attributed to a single electronic component (in one 
active Hubbard band) experiencing spin-fluctuation correlations that reduce 
its spin contribution. In Fe superconductors, no such correlation effects 
are required. 

We extend these considerations to the NMR data obtained for other 
iron-based superconductors. Our results are almost identical to those for 
K$_y$Fe$_{2-x}$Se$_2$,\cite{YuW_11011017, ImaiT_11014967} while the binary 
iron selenides also show a similarly substantial
$T^2$-type increase up to 300 K on top of a constant contribution [shown in Fig.~\ref{kn3} (b)
using $^{77}K$ data adapted from Ref.~[\onlinecite{Imai_prl_102_177005}]. 
This indicates universal behavior in the FeSe planes of the chalcogenides.
In the three known iron pnictide superconductors, weak 
increases of $K(T)$ above $T_c$ have also been interpreted as pseudogap 
behavior. In both the 1111\cite{Imai_08044026, Nakai_JPSJ} and 
111\cite{Jeglic_LiFeAs_NMR, Li_JPSJ_79} systems, the $T$-dependent part 
of $K$ and $1/T_1T$ is small and almost linear, suggesting only a weak 
contribution from quasi-2D spin fluctuations. The 122 systems deviate from
this paradigm only at low temperatures: \cite{Ning_PRL_104} while $K$ has a small spin-fluctuation 
part with a possibly quadratic (3D) $T$-dependence, $1/T_1T$ shows an increase 
towards low $T$ for most dopings, indicating a role for magnetic modes with 
${\bf Q} \ne 0$.

Thus the two-component framework forms a basis for the normal-state NMR 
response of all the Fe superconductors. This general result provides 
important guidance to both experimental and theoretical understanding of 
these materials far beyond NMR measurements. The importance of local 
spin fluctuations and the lack of relevance of particular magnetic modes, 
despite the proximity to magnetic phases, are essential clues to the 
mechanism for superconductivity. The clear presence of both Fermi-liquid 
and local-moment behavior reflects the generic property of the Fe 
superconductors that both conduction and valence electrons play a crucial 
role.

In summary, our NMR measurements on the intercalated iron selenide 
superconductor (Tl,Rb)$_{y}$Fe$_{2-x}$Se$_2$ show a large increase of 
the isotropic Knight shift and the spin-lattice relaxation rate with 
temperature above $T_c$. The temperature-dependence of this contribution 
is quadratic. We deduce that pseudogap-like behavior in the paramagnetic, 
metallic phase is the consequence of two separate contributions, one from 
itinerant electrons, which become superconducting below $T_c$, and one from 
local 3D spin fluctuations of disordered Fe moments, which become dominant 
above $T_c$. Such a two-component model provides a framework for understanding 
the $K$ and $1/T_1 T$ data in all Fe superconductors. 

The authors acknowledge W. Bao, S. E. Brown, J. Hu, T. Li, Z. Y. Lu, Y. Su, 
X. Q. Wang, T. Xiang, D.-X. Yao, G. Zhang, Q. M. Zhang, and X. J. Zhou for 
helpful discussions. This work was supported by the NSFC under Grant 
Nos.~10974254, 11074304, and 10874244, and the National Basic Research 
Program of China under Grant Nos.~2010CB923000 and 2011CBA00112. W.Y. is 
supported by the ``New Century Program for Excellent Talent in Universities'' 
of the Chinese MoE.
 

\begin{thebibliography}{46}
\expandafter\ifx\csname natexlab\endcsname\relax\def\natexlab#1{#1}\fi
\expandafter\ifx\csname bibnamefont\endcsname\relax
  \def\bibnamefont#1{#1}\fi
\expandafter\ifx\csname bibfnamefont\endcsname\relax
  \def\bibfnamefont#1{#1}\fi
\expandafter\ifx\csname citenamefont\endcsname\relax
  \def\citenamefont#1{#1}\fi
\expandafter\ifx\csname url\endcsname\relax
  \def\url#1{\texttt{#1}}\fi
\expandafter\ifx\csname urlprefix\endcsname\relax\def\urlprefix{URL }\fi
\providecommand{\bibinfo}[2]{#2}
\providecommand{\eprint}[2][]{\url{#2}}

\bibitem[{\citenamefont{Warren et~al.}(1989)}]{Warren_PG}
\bibinfo{author}{\bibfnamefont{W.~W.} \bibnamefont{Warren}}
  \bibnamefont{et~al.}, \bibinfo{journal}{Phys. Rev. Lett.}
  \textbf{\bibinfo{volume}{62}}, \bibinfo{pages}{1193} (\bibinfo{year}{1989}).

\bibitem[{\citenamefont{Alloul et~al.}(1989)\citenamefont{Alloul, Ohno, and
  Mendels}}]{Alloul_PG}
\bibinfo{author}{\bibfnamefont{H.}~\bibnamefont{Alloul}},
  \bibinfo{author}{\bibfnamefont{T.}~\bibnamefont{Ohno}}, \bibnamefont{and}
  \bibinfo{author}{\bibfnamefont{P.}~\bibnamefont{Mendels}},
  \bibinfo{journal}{Phys. Rev. Lett.} \textbf{\bibinfo{volume}{63}},
  \bibinfo{pages}{1700} (\bibinfo{year}{1989}).

\bibitem[{\citenamefont{Kamihara et~al.}(2008)\citenamefont{Kamihara, Watanabe,
  Hirano, and Hosono}}]{Hosono_Jacs_130_3296}
\bibinfo{author}{\bibfnamefont{Y.}~\bibnamefont{Kamihara}},
  \bibinfo{author}{\bibfnamefont{T.}~\bibnamefont{Watanabe}},
  \bibinfo{author}{\bibfnamefont{M.}~\bibnamefont{Hirano}}, \bibnamefont{and}
  \bibinfo{author}{\bibfnamefont{H.}~\bibnamefont{Hosono}},
  \bibinfo{journal}{J. Am. Chem. Soc.} \textbf{\bibinfo{volume}{130}},
  \bibinfo{pages}{3296} (\bibinfo{year}{2008}).

\bibitem[{\citenamefont{Chen et~al.}(2008{\natexlab{a}})\citenamefont{Chen, Wu,
  Wu, Liu, Chen, and Fang}}]{Chen_Nature_453_761}
\bibinfo{author}{\bibfnamefont{X.~H.} \bibnamefont{Chen}},
  \bibinfo{author}{\bibfnamefont{T.}~\bibnamefont{Wu}},
  \bibinfo{author}{\bibfnamefont{G.}~\bibnamefont{Wu}},
  \bibinfo{author}{\bibfnamefont{R.~H.} \bibnamefont{Liu}},
  \bibinfo{author}{\bibfnamefont{H.}~\bibnamefont{Chen}}, \bibnamefont{and}
  \bibinfo{author}{\bibfnamefont{D.~F.} \bibnamefont{Fang}},
  \bibinfo{journal}{Nature} \textbf{\bibinfo{volume}{453}},
  \bibinfo{pages}{761} (\bibinfo{year}{2008}{\natexlab{a}}).

\bibitem[{\citenamefont{Ren et~al.}(2008)\citenamefont{Ren, Lu, Yang, Yi, Shen,
  Li, Che, Dong, Sun, Zhou et~al.}}]{Ren_CPL_12_105}
\bibinfo{author}{\bibfnamefont{Z.~A.} \bibnamefont{Ren}},
  \bibinfo{author}{\bibfnamefont{W.}~\bibnamefont{Lu}},
  \bibinfo{author}{\bibfnamefont{J.}~\bibnamefont{Yang}},
  \bibinfo{author}{\bibfnamefont{W.}~\bibnamefont{Yi}},
  \bibinfo{author}{\bibfnamefont{X.~L.} \bibnamefont{Shen}},
  \bibinfo{author}{\bibfnamefont{Z.~C.} \bibnamefont{Li}},
  \bibinfo{author}{\bibfnamefont{G.~C.} \bibnamefont{Che}},
  \bibinfo{author}{\bibfnamefont{X.~L.} \bibnamefont{Dong}},
  \bibinfo{author}{\bibfnamefont{L.~L.} \bibnamefont{Sun}},
  \bibinfo{author}{\bibfnamefont{F.}~\bibnamefont{Zhou}}, \bibnamefont{et~al.},
  \bibinfo{journal}{Chinese Phys. Lett.} \textbf{\bibinfo{volume}{25}},
  \bibinfo{pages}{2215} (\bibinfo{year}{2008}).

\bibitem[{\citenamefont{Chen et~al.}(2008{\natexlab{b}})\citenamefont{Chen, Li,
  Wu, Li, Hu, Dong, Zheng, Luo, and Wang}}]{Chen_PRL_100_247002}
\bibinfo{author}{\bibfnamefont{G.~F.} \bibnamefont{Chen}},
  \bibinfo{author}{\bibfnamefont{Z.}~\bibnamefont{Li}},
  \bibinfo{author}{\bibfnamefont{D.}~\bibnamefont{Wu}},
  \bibinfo{author}{\bibfnamefont{G.}~\bibnamefont{Li}},
  \bibinfo{author}{\bibfnamefont{W.~Z.} \bibnamefont{Hu}},
  \bibinfo{author}{\bibfnamefont{J.}~\bibnamefont{Dong}},
  \bibinfo{author}{\bibfnamefont{P.}~\bibnamefont{Zheng}},
  \bibinfo{author}{\bibfnamefont{J.~L.} \bibnamefont{Luo}}, \bibnamefont{and}
  \bibinfo{author}{\bibfnamefont{N.~L.} \bibnamefont{Wang}},
  \bibinfo{journal}{Phys. Rev. Lett.} \textbf{\bibinfo{volume}{100}},
  \bibinfo{pages}{247002} (\bibinfo{year}{2008}{\natexlab{b}}).

\bibitem[{\citenamefont{Wu et~al.}(2008{\natexlab{a}})\citenamefont{Wu, Chen,
  Wu, Xie, Yan, Liu, Wang, Ying, and Chen}}]{ChenXHSUS}
\bibinfo{author}{\bibfnamefont{G.}~\bibnamefont{Wu}},
  \bibinfo{author}{\bibfnamefont{H.}~\bibnamefont{Chen}},
  \bibinfo{author}{\bibfnamefont{T.}~\bibnamefont{Wu}},
  \bibinfo{author}{\bibfnamefont{Y.~L.} \bibnamefont{Xie}},
  \bibinfo{author}{\bibfnamefont{Y.~J.} \bibnamefont{Yan}},
  \bibinfo{author}{\bibfnamefont{R.~H.} \bibnamefont{Liu}},
  \bibinfo{author}{\bibfnamefont{X.~F.} \bibnamefont{Wang}},
  \bibinfo{author}{\bibfnamefont{J.~J.} \bibnamefont{Ying}}, \bibnamefont{and}
  \bibinfo{author}{\bibfnamefont{X.~H.} \bibnamefont{Chen}},
  \bibinfo{journal}{Condensed Matter} \textbf{\bibinfo{volume}{20}},
  \bibinfo{pages}{422201} (\bibinfo{year}{2008}{\natexlab{a}}).

\bibitem[{\citenamefont{Yan et~al.}(2008)\citenamefont{Yan, Kreyssig, Nandi,
  Ni, Bud'ko, Kracher, McQueeney, McCallum, Lograsso, Goldman
  et~al.}}]{CanfieldSUS}
\bibinfo{author}{\bibfnamefont{J.~Q.} \bibnamefont{Yan}},
  \bibinfo{author}{\bibfnamefont{A.}~\bibnamefont{Kreyssig}},
  \bibinfo{author}{\bibfnamefont{S.}~\bibnamefont{Nandi}},
  \bibinfo{author}{\bibfnamefont{N.}~\bibnamefont{Ni}},
  \bibinfo{author}{\bibfnamefont{S.~L.} \bibnamefont{Bud'ko}},
  \bibinfo{author}{\bibfnamefont{A.}~\bibnamefont{Kracher}},
  \bibinfo{author}{\bibfnamefont{R.~J.} \bibnamefont{McQueeney}},
  \bibinfo{author}{\bibfnamefont{R.~W.} \bibnamefont{McCallum}},
  \bibinfo{author}{\bibfnamefont{T.~A.} \bibnamefont{Lograsso}},
  \bibinfo{author}{\bibfnamefont{A.~I.} \bibnamefont{Goldman}},
  \bibnamefont{et~al.}, \bibinfo{journal}{Phys. Rev. B}
  \textbf{\bibinfo{volume}{78}}, \bibinfo{pages}{024516}
  (\bibinfo{year}{2008}).

\bibitem[{\citenamefont{Klingeler et~al.}(2010)\citenamefont{Klingeler, Leps,
  Hellmann, Popa, Stockert, Hess, Kataev, Grafe, Hammerath, Lang
  et~al.}}]{BuechnerSUS}
\bibinfo{author}{\bibfnamefont{R.}~\bibnamefont{Klingeler}},
  \bibinfo{author}{\bibfnamefont{N.}~\bibnamefont{Leps}},
  \bibinfo{author}{\bibfnamefont{I.}~\bibnamefont{Hellmann}},
  \bibinfo{author}{\bibfnamefont{A.}~\bibnamefont{Popa}},
  \bibinfo{author}{\bibfnamefont{U.}~\bibnamefont{Stockert}},
  \bibinfo{author}{\bibfnamefont{C.}~\bibnamefont{Hess}},
  \bibinfo{author}{\bibfnamefont{V.}~\bibnamefont{Kataev}},
  \bibinfo{author}{\bibfnamefont{H.~J.} \bibnamefont{Grafe}},
  \bibinfo{author}{\bibfnamefont{F.}~\bibnamefont{Hammerath}},
  \bibinfo{author}{\bibfnamefont{G.}~\bibnamefont{Lang}}, \bibnamefont{et~al.},
  \bibinfo{journal}{Phys. Rev. B} \textbf{\bibinfo{volume}{81}},
  \bibinfo{pages}{024506} (\bibinfo{year}{2010}).

\bibitem[{\citenamefont{Ning et~al.}(2010)\citenamefont{Ning, Ahilan, Imai,
  Sefat, McGuire, Sales, Mandrus, Cheng, Shen, and Wen}}]{Ning_PRL_104}
\bibinfo{author}{\bibfnamefont{F.~L.} \bibnamefont{Ning}},
  \bibinfo{author}{\bibfnamefont{K.}~\bibnamefont{Ahilan}},
  \bibinfo{author}{\bibfnamefont{T.}~\bibnamefont{Imai}},
  \bibinfo{author}{\bibfnamefont{A.~S.} \bibnamefont{Sefat}},
  \bibinfo{author}{\bibfnamefont{M.~A.} \bibnamefont{McGuire}},
  \bibinfo{author}{\bibfnamefont{B.~C.} \bibnamefont{Sales}},
  \bibinfo{author}{\bibfnamefont{D.}~\bibnamefont{Mandrus}},
  \bibinfo{author}{\bibfnamefont{P.}~\bibnamefont{Cheng}},
  \bibinfo{author}{\bibfnamefont{B.}~\bibnamefont{Shen}}, \bibnamefont{and}
  \bibinfo{author}{\bibfnamefont{H.~H.} \bibnamefont{Wen}},
  \bibinfo{journal}{Phys. Rev. Lett.} \textbf{\bibinfo{volume}{104}},
  \bibinfo{pages}{037001} (\bibinfo{year}{2010}).

\bibitem[{\citenamefont{Nakai et~al.}(2008)\citenamefont{Nakai, Ishida,
  Kmihara, Hirano, and Hosono}}]{Nakai_JPSJ}
\bibinfo{author}{\bibfnamefont{Y.}~\bibnamefont{Nakai}},
  \bibinfo{author}{\bibfnamefont{K.}~\bibnamefont{Ishida}},
  \bibinfo{author}{\bibfnamefont{Y.}~\bibnamefont{Kmihara}},
  \bibinfo{author}{\bibfnamefont{M.}~\bibnamefont{Hirano}}, \bibnamefont{and}
  \bibinfo{author}{\bibfnamefont{H.}~\bibnamefont{Hosono}},
  \bibinfo{journal}{J. Phy. Soc. Jpn.} \textbf{\bibinfo{volume}{77}},
  \bibinfo{pages}{073701} (\bibinfo{year}{2008}).

\bibitem[{\citenamefont{Guo et~al.}(2010)\citenamefont{Guo, Jin, Wang, Wang,
  Zhu, Zhou, He, and Chen}}]{Guo_PRB_82_180520}
\bibinfo{author}{\bibfnamefont{J.}~\bibnamefont{Guo}},
  \bibinfo{author}{\bibfnamefont{S.}~\bibnamefont{Jin}},
  \bibinfo{author}{\bibfnamefont{G.}~\bibnamefont{Wang}},
  \bibinfo{author}{\bibfnamefont{S.}~\bibnamefont{Wang}},
  \bibinfo{author}{\bibfnamefont{K.}~\bibnamefont{Zhu}},
  \bibinfo{author}{\bibfnamefont{T.}~\bibnamefont{Zhou}},
  \bibinfo{author}{\bibfnamefont{M.}~\bibnamefont{He}}, \bibnamefont{and}
  \bibinfo{author}{\bibfnamefont{X.}~\bibnamefont{Chen}},
  \bibinfo{journal}{Phys. Rev. B} \textbf{\bibinfo{volume}{82}},
  \bibinfo{pages}{180520(R)} (\bibinfo{year}{2010}).

\bibitem[{\citenamefont{Mizuguchi et~al.}(2011)\citenamefont{Mizuguchi, Takeya,
  Kawasaki, Ozaki, Tsuda, Yamaguchi, and Takano}}]{Mizuguchi_10124950}
\bibinfo{author}{\bibfnamefont{Y.}~\bibnamefont{Mizuguchi}},
  \bibinfo{author}{\bibfnamefont{H.}~\bibnamefont{Takeya}},
  \bibinfo{author}{\bibfnamefont{Y.}~\bibnamefont{Kawasaki}},
  \bibinfo{author}{\bibfnamefont{T.}~\bibnamefont{Ozaki}},
  \bibinfo{author}{\bibfnamefont{S.}~\bibnamefont{Tsuda}},
  \bibinfo{author}{\bibfnamefont{T.}~\bibnamefont{Yamaguchi}},
  \bibnamefont{and} \bibinfo{author}{\bibfnamefont{Y.}~\bibnamefont{Takano}},
  \bibinfo{journal}{Appl. Phys. Lett.} \textbf{\bibinfo{volume}{98}},
  \bibinfo{pages}{042511} (\bibinfo{year}{2011}).

\bibitem[{\citenamefont{Wang et~al.}(2011{\natexlab{a}})\citenamefont{Wang,
  Ying, Yan, Liu, Luo, Li, Wang, Zhang, Ye, Cheng et~al.}}]{Chen_CM_10125525}
\bibinfo{author}{\bibfnamefont{A.~F.} \bibnamefont{Wang}},
  \bibinfo{author}{\bibfnamefont{J.~J.} \bibnamefont{Ying}},
  \bibinfo{author}{\bibfnamefont{Y.~J.} \bibnamefont{Yan}},
  \bibinfo{author}{\bibfnamefont{R.~H.} \bibnamefont{Liu}},
  \bibinfo{author}{\bibfnamefont{X.~G.} \bibnamefont{Luo}},
  \bibinfo{author}{\bibfnamefont{Z.~Y.} \bibnamefont{Li}},
  \bibinfo{author}{\bibfnamefont{X.~F.} \bibnamefont{Wang}},
  \bibinfo{author}{\bibfnamefont{M.}~\bibnamefont{Zhang}},
  \bibinfo{author}{\bibfnamefont{G.~J.} \bibnamefont{Ye}},
  \bibinfo{author}{\bibfnamefont{P.}~\bibnamefont{Cheng}},
  \bibnamefont{et~al.}, \bibinfo{journal}{Phys. Rev. B}
  \textbf{\bibinfo{volume}{83}}, \bibinfo{pages}{060512}
  (\bibinfo{year}{2011}{\natexlab{a}}).

\bibitem[{\citenamefont{Fang et~al.}(2010)\citenamefont{Fang, Wang, Dong, Li,
  Feng, Chen, and Yuan}}]{FangMH_CM_10125236}
\bibinfo{author}{\bibfnamefont{M.}~\bibnamefont{Fang}},
  \bibinfo{author}{\bibfnamefont{H.}~\bibnamefont{Wang}},
  \bibinfo{author}{\bibfnamefont{C.}~\bibnamefont{Dong}},
  \bibinfo{author}{\bibfnamefont{Z.}~\bibnamefont{Li}},
  \bibinfo{author}{\bibfnamefont{C.}~\bibnamefont{Feng}},
  \bibinfo{author}{\bibfnamefont{J.}~\bibnamefont{Chen}}, \bibnamefont{and}
  \bibinfo{author}{\bibfnamefont{H.~Q.} \bibnamefont{Yuan}},
  \bibinfo{journal}{Europhys. Lett.} \textbf{\bibinfo{volume}{94}},
  \bibinfo{pages}{27009} (\bibinfo{year}{2010}).

\bibitem[{\citenamefont{Wang et~al.}(2011{\natexlab{b}})\citenamefont{Wang, He,
  Xia, and Chen}}]{Chen_CM_11010789}
\bibinfo{author}{\bibfnamefont{D.~M.} \bibnamefont{Wang}},
  \bibinfo{author}{\bibfnamefont{J.~B.} \bibnamefont{He}},
  \bibinfo{author}{\bibfnamefont{T.-L.} \bibnamefont{Xia}}, \bibnamefont{and}
  \bibinfo{author}{\bibfnamefont{G.~F.} \bibnamefont{Chen}},
  \bibinfo{journal}{Phys. Rev. B} \textbf{\bibinfo{volume}{83}},
  \bibinfo{pages}{132502} (\bibinfo{year}{2011}{\natexlab{b}}).

\bibitem[{\citenamefont{Zavalij et~al.}(2011)\citenamefont{Zavalij, Bao, Wang,
  Ying, Chen, Wang, He, Wang, Chen, Hsieh et~al.}}]{BaoW_11014882}
\bibinfo{author}{\bibfnamefont{P.}~\bibnamefont{Zavalij}},
  \bibinfo{author}{\bibfnamefont{W.}~\bibnamefont{Bao}},
  \bibinfo{author}{\bibfnamefont{X.~F.} \bibnamefont{Wang}},
  \bibinfo{author}{\bibfnamefont{J.~J.} \bibnamefont{Ying}},
  \bibinfo{author}{\bibfnamefont{X.~H.} \bibnamefont{Chen}},
  \bibinfo{author}{\bibfnamefont{D.~M.} \bibnamefont{Wang}},
  \bibinfo{author}{\bibfnamefont{J.~B.} \bibnamefont{He}},
  \bibinfo{author}{\bibfnamefont{X.~Q.} \bibnamefont{Wang}},
  \bibinfo{author}{\bibfnamefont{G.~F.} \bibnamefont{Chen}},
  \bibinfo{author}{\bibfnamefont{P.~Y.} \bibnamefont{Hsieh}},
  \bibnamefont{et~al.}, \bibinfo{journal}{Phys. Rev. B}
  \textbf{\bibinfo{volume}{83}}, \bibinfo{pages}{132509}
  (\bibinfo{year}{2011}).

\bibitem[{\citenamefont{Ye et~al.}(2011)\citenamefont{Ye, Chi, Bao, Wang, Ying,
  Chen, Wang, Dong, and Fang}}]{Ye_11022882}
\bibinfo{author}{\bibfnamefont{F.}~\bibnamefont{Ye}},
  \bibinfo{author}{\bibfnamefont{S.}~\bibnamefont{Chi}},
  \bibinfo{author}{\bibfnamefont{W.}~\bibnamefont{Bao}},
  \bibinfo{author}{\bibfnamefont{X.~F.} \bibnamefont{Wang}},
  \bibinfo{author}{\bibfnamefont{J.~J.} \bibnamefont{Ying}},
  \bibinfo{author}{\bibfnamefont{X.~H.} \bibnamefont{Chen}},
  \bibinfo{author}{\bibfnamefont{H.~D.} \bibnamefont{Wang}},
  \bibinfo{author}{\bibfnamefont{C.~H.} \bibnamefont{Dong}}, \bibnamefont{and}
  \bibinfo{author}{\bibfnamefont{M.}~\bibnamefont{Fang}},
  \bibinfo{journal}{Phys. Rev. Lett.} \textbf{\bibinfo{volume}{107}},
  \bibinfo{pages}{137003} (\bibinfo{year}{2011}).

\bibitem[{\citenamefont{Bao et~al.}(2011)\citenamefont{Bao, Huang, Chen, Green,
  Wang, He, Wang, and Qiu}}]{Bao_11020830}
\bibinfo{author}{\bibfnamefont{W.}~\bibnamefont{Bao}},
  \bibinfo{author}{\bibfnamefont{Q.}~\bibnamefont{Huang}},
  \bibinfo{author}{\bibfnamefont{G.~F.} \bibnamefont{Chen}},
  \bibinfo{author}{\bibfnamefont{M.~A.} \bibnamefont{Green}},
  \bibinfo{author}{\bibfnamefont{D.~M.} \bibnamefont{Wang}},
  \bibinfo{author}{\bibfnamefont{J.~B.} \bibnamefont{He}},
  \bibinfo{author}{\bibfnamefont{X.~Q.} \bibnamefont{Wang}}, \bibnamefont{and}
  \bibinfo{author}{\bibfnamefont{Y.}~\bibnamefont{Qiu}},
  \bibinfo{journal}{Chinese Phys. Lett.} \textbf{\bibinfo{volume}{28}},
  \bibinfo{pages}{086104} (\bibinfo{year}{2011}).

\bibitem[{\citenamefont{Liu et~al.}(2011)\citenamefont{Liu, Luo, Zhang, Wang,
  Ying, Wang, Yan, Xiang, Cheng, Ye et~al.}}]{ChenXH_11022783}
\bibinfo{author}{\bibfnamefont{R.~H.} \bibnamefont{Liu}},
  \bibinfo{author}{\bibfnamefont{X.~G.} \bibnamefont{Luo}},
  \bibinfo{author}{\bibfnamefont{M.}~\bibnamefont{Zhang}},
  \bibinfo{author}{\bibfnamefont{A.~F.} \bibnamefont{Wang}},
  \bibinfo{author}{\bibfnamefont{J.~J.} \bibnamefont{Ying}},
  \bibinfo{author}{\bibfnamefont{X.~F.} \bibnamefont{Wang}},
  \bibinfo{author}{\bibfnamefont{Y.~J.} \bibnamefont{Yan}},
  \bibinfo{author}{\bibfnamefont{Z.~J.} \bibnamefont{Xiang}},
  \bibinfo{author}{\bibfnamefont{P.}~\bibnamefont{Cheng}},
  \bibinfo{author}{\bibfnamefont{G.~J.} \bibnamefont{Ye}},
  \bibnamefont{et~al.}, \bibinfo{journal}{Europhys. Lett.}
  \textbf{\bibinfo{volume}{94}}, \bibinfo{pages}{27008} (\bibinfo{year}{2011}).

\bibitem[{\citenamefont{Shermadini et~al.}(2011)\citenamefont{Shermadini,
  Krzton-Maziopa, Bendele, Khasanov, Luetkens, Conder, Pomjakushina, Weyeneth,
  Pomjakushin, Bossen et~al.}}]{Shermadini_11011873}
\bibinfo{author}{\bibfnamefont{Z.}~\bibnamefont{Shermadini}},
  \bibinfo{author}{\bibfnamefont{A.}~\bibnamefont{Krzton-Maziopa}},
  \bibinfo{author}{\bibfnamefont{M.}~\bibnamefont{Bendele}},
  \bibinfo{author}{\bibfnamefont{R.}~\bibnamefont{Khasanov}},
  \bibinfo{author}{\bibfnamefont{H.}~\bibnamefont{Luetkens}},
  \bibinfo{author}{\bibfnamefont{K.}~\bibnamefont{Conder}},
  \bibinfo{author}{\bibfnamefont{E.}~\bibnamefont{Pomjakushina}},
  \bibinfo{author}{\bibfnamefont{S.}~\bibnamefont{Weyeneth}},
  \bibinfo{author}{\bibfnamefont{V.}~\bibnamefont{Pomjakushin}},
  \bibinfo{author}{\bibfnamefont{O.}~\bibnamefont{Bossen}},
  \bibnamefont{et~al.}, \bibinfo{journal}{Phys. Rev. Lett.}
  \textbf{\bibinfo{volume}{106}}, \bibinfo{pages}{117602}
  (\bibinfo{year}{2011}).

\bibitem[{\citenamefont{Yu et~al.}(2011)\citenamefont{Yu, Ma, He, Wang, Xia,
  Chen, and Bao}}]{YuW_11011017}
\bibinfo{author}{\bibfnamefont{W.}~\bibnamefont{Yu}},
  \bibinfo{author}{\bibfnamefont{L.}~\bibnamefont{Ma}},
  \bibinfo{author}{\bibfnamefont{J.~B.} \bibnamefont{He}},
  \bibinfo{author}{\bibfnamefont{D.~M.} \bibnamefont{Wang}},
  \bibinfo{author}{\bibfnamefont{T.~L.} \bibnamefont{Xia}},
  \bibinfo{author}{\bibfnamefont{G.~F.} \bibnamefont{Chen}}, \bibnamefont{and}
  \bibinfo{author}{\bibfnamefont{W.}~\bibnamefont{Bao}},
  \bibinfo{journal}{Phys. Rev. Lett.} \textbf{\bibinfo{volume}{106}},
  \bibinfo{pages}{197001} (\bibinfo{year}{2011}).

\bibitem[{\citenamefont{Torchetti et~al.}(2011)\citenamefont{Torchetti, Fu,
  Christensen, Nelson, Imai, Lei, and Petrovic}}]{ImaiT_11014967}
\bibinfo{author}{\bibfnamefont{D.~A.} \bibnamefont{Torchetti}},
  \bibinfo{author}{\bibfnamefont{M.}~\bibnamefont{Fu}},
  \bibinfo{author}{\bibfnamefont{D.~C.} \bibnamefont{Christensen}},
  \bibinfo{author}{\bibfnamefont{K.~J.} \bibnamefont{Nelson}},
  \bibinfo{author}{\bibfnamefont{T.}~\bibnamefont{Imai}},
  \bibinfo{author}{\bibfnamefont{H.~C.} \bibnamefont{Lei}}, \bibnamefont{and}
  \bibinfo{author}{\bibfnamefont{C.}~\bibnamefont{Petrovic}},
  \bibinfo{journal}{Phys. Rev. B} \textbf{\bibinfo{volume}{83}},
  \bibinfo{pages}{104508} (\bibinfo{year}{2011}).

\bibitem[{\citenamefont{Ma et~al.}(2011{\natexlab{a}})\citenamefont{Ma, Ji,
  Zhang, He, Wang, Chen, Bao, and Yu}}]{Ma11023888}
\bibinfo{author}{\bibfnamefont{L.}~\bibnamefont{Ma}},
  \bibinfo{author}{\bibfnamefont{G.~F.} \bibnamefont{Ji}},
  \bibinfo{author}{\bibfnamefont{J.}~\bibnamefont{Zhang}},
  \bibinfo{author}{\bibfnamefont{J.~B.} \bibnamefont{He}},
  \bibinfo{author}{\bibfnamefont{D.~M.} \bibnamefont{Wang}},
  \bibinfo{author}{\bibfnamefont{G.~F.} \bibnamefont{Chen}},
  \bibinfo{author}{\bibfnamefont{W.}~\bibnamefont{Bao}}, \bibnamefont{and}
  \bibinfo{author}{\bibfnamefont{W.}~\bibnamefont{Yu}}, \bibinfo{journal}{Phys.
  Rev. B} \textbf{\bibinfo{volume}{83}}, \bibinfo{pages}{174510}
  (\bibinfo{year}{2011}{\natexlab{a}}).

\bibitem[{\citenamefont{Zhang et~al.}(2011{\natexlab{a}})\citenamefont{Zhang,
  Yang, Xu, Ye, Chen, He, Jiang, Xie, Ying, Wang et~al.}}]{Feng_CM_10125980}
\bibinfo{author}{\bibfnamefont{Y.}~\bibnamefont{Zhang}},
  \bibinfo{author}{\bibfnamefont{L.~X.} \bibnamefont{Yang}},
  \bibinfo{author}{\bibfnamefont{M.}~\bibnamefont{Xu}},
  \bibinfo{author}{\bibfnamefont{Z.~R.} \bibnamefont{Ye}},
  \bibinfo{author}{\bibfnamefont{F.}~\bibnamefont{Chen}},
  \bibinfo{author}{\bibfnamefont{C.}~\bibnamefont{He}},
  \bibinfo{author}{\bibfnamefont{J.}~\bibnamefont{Jiang}},
  \bibinfo{author}{\bibfnamefont{B.~P.} \bibnamefont{Xie}},
  \bibinfo{author}{\bibfnamefont{J.~J.} \bibnamefont{Ying}},
  \bibinfo{author}{\bibfnamefont{X.~F.} \bibnamefont{Wang}},
  \bibnamefont{et~al.}, \bibinfo{journal}{Nature Materials}
  \textbf{\bibinfo{volume}{10}}, \bibinfo{pages}{273}
  (\bibinfo{year}{2011}{\natexlab{a}}).

\bibitem[{\citenamefont{Mou et~al.}(2011)\citenamefont{Mou, Liu, Jia, He, Peng,
  Zhao, Yu, Liu, He, Dong et~al.}}]{ZhouXJ_11014556}
\bibinfo{author}{\bibfnamefont{D.}~\bibnamefont{Mou}},
  \bibinfo{author}{\bibfnamefont{S.}~\bibnamefont{Liu}},
  \bibinfo{author}{\bibfnamefont{X.}~\bibnamefont{Jia}},
  \bibinfo{author}{\bibfnamefont{J.}~\bibnamefont{He}},
  \bibinfo{author}{\bibfnamefont{Y.}~\bibnamefont{Peng}},
  \bibinfo{author}{\bibfnamefont{L.}~\bibnamefont{Zhao}},
  \bibinfo{author}{\bibfnamefont{L.}~\bibnamefont{Yu}},
  \bibinfo{author}{\bibfnamefont{G.}~\bibnamefont{Liu}},
  \bibinfo{author}{\bibfnamefont{S.}~\bibnamefont{He}},
  \bibinfo{author}{\bibfnamefont{X.}~\bibnamefont{Dong}}, \bibnamefont{et~al.},
  \bibinfo{journal}{Phys. Rev. Lett.} \textbf{\bibinfo{volume}{106}},
  \bibinfo{pages}{107001} (\bibinfo{year}{2011}).

\bibitem[{\citenamefont{Qian et~al.}(2011)\citenamefont{Qian, Wang, Jin, Zhang,
  Richard, Xu, Dai, Fang, Guo, Chen et~al.}}]{Ding_10126017}
\bibinfo{author}{\bibfnamefont{T.}~\bibnamefont{Qian}},
  \bibinfo{author}{\bibfnamefont{X.~P.} \bibnamefont{Wang}},
  \bibinfo{author}{\bibfnamefont{W.~C.} \bibnamefont{Jin}},
  \bibinfo{author}{\bibfnamefont{P.}~\bibnamefont{Zhang}},
  \bibinfo{author}{\bibfnamefont{P.}~\bibnamefont{Richard}},
  \bibinfo{author}{\bibfnamefont{G.}~\bibnamefont{Xu}},
  \bibinfo{author}{\bibfnamefont{X.}~\bibnamefont{Dai}},
  \bibinfo{author}{\bibfnamefont{Z.}~\bibnamefont{Fang}},
  \bibinfo{author}{\bibfnamefont{J.~G.} \bibnamefont{Guo}},
  \bibinfo{author}{\bibfnamefont{X.~L.} \bibnamefont{Chen}},
  \bibnamefont{et~al.}, \bibinfo{journal}{Phys. Rev. Lett.}
  \textbf{\bibinfo{volume}{106}}, \bibinfo{pages}{187001}
  (\bibinfo{year}{2011}).

\bibitem[{Me2()}]{Me2Se}
\bibinfo{note}{Me2Se was used as the reference material.}

\bibitem[{\citenamefont{Bankay et~al.}(1994)\citenamefont{Bankay, Mali, Roos,
  and Brinkmann}}]{Bankay_PRB_50_6416}
\bibinfo{author}{\bibfnamefont{M.}~\bibnamefont{Bankay}},
  \bibinfo{author}{\bibfnamefont{M.}~\bibnamefont{Mali}},
  \bibinfo{author}{\bibfnamefont{J.}~\bibnamefont{Roos}}, \bibnamefont{and}
  \bibinfo{author}{\bibfnamefont{D.}~\bibnamefont{Brinkmann}},
  \bibinfo{journal}{Phys. Rev. B} \textbf{\bibinfo{volume}{50}},
  \bibinfo{pages}{6416} (\bibinfo{year}{1994}).

\bibitem[{\citenamefont{Walstedt}(2008)}]{Walstedybook}
\bibinfo{author}{\bibfnamefont{R.}~\bibnamefont{Walstedt}},
  \emph{\bibinfo{title}{The NMR Probe of High-Tc Materials}}
  (\bibinfo{publisher}{Springer, Berlin, Heidelberg}, \bibinfo{year}{2008}).

\bibitem[{\citenamefont{Imai et~al.}(2009)\citenamefont{Imai, Ahilan, Ning,
  McQueen, and Cava}}]{Imai_prl_102_177005}
\bibinfo{author}{\bibfnamefont{T.}~\bibnamefont{Imai}},
  \bibinfo{author}{\bibfnamefont{K.}~\bibnamefont{Ahilan}},
  \bibinfo{author}{\bibfnamefont{F.~L.} \bibnamefont{Ning}},
  \bibinfo{author}{\bibfnamefont{T.~M.} \bibnamefont{McQueen}},
  \bibnamefont{and} \bibinfo{author}{\bibfnamefont{R.~J.} \bibnamefont{Cava}},
  \bibinfo{journal}{Phys. Rev. Lett.} \textbf{\bibinfo{volume}{102}},
  \bibinfo{pages}{177005} (\bibinfo{year}{2009}).

\bibitem[{\citenamefont{Slichter}(1990)}]{slichterbook}
\bibinfo{author}{\bibfnamefont{C.~P.} \bibnamefont{Slichter}},
  \emph{\bibinfo{title}{Principles of Magnetic Resonance}}
  (\bibinfo{publisher}{Springer-Verlag}, \bibinfo{year}{1990}),
  \bibinfo{edition}{3rd} ed.

\bibitem[{\citenamefont{Abragam}(1994)}]{Abragambook}
\bibinfo{author}{\bibfnamefont{A.}~\bibnamefont{Abragam}},
  \emph{\bibinfo{title}{The Principles of Nuclear Magnetism}}
  (\bibinfo{publisher}{Oxford University Press, London}, \bibinfo{year}{1994}).

\bibitem[{\citenamefont{Ma et~al.}(2011{\natexlab{b}})\citenamefont{Ma, Chen,
  Yao, Zhang, Zhang, Xia, and Yu}}]{MaNaFeAs}
\bibinfo{author}{\bibfnamefont{L.}~\bibnamefont{Ma}},
  \bibinfo{author}{\bibfnamefont{G.~F.} \bibnamefont{Chen}},
  \bibinfo{author}{\bibfnamefont{D.~X.} \bibnamefont{Yao}},
  \bibinfo{author}{\bibfnamefont{J.}~\bibnamefont{Zhang}},
  \bibinfo{author}{\bibfnamefont{S.}~\bibnamefont{Zhang}},
  \bibinfo{author}{\bibfnamefont{T.~L.} \bibnamefont{Xia}}, \bibnamefont{and}
  \bibinfo{author}{\bibfnamefont{W.}~\bibnamefont{Yu}}, \bibinfo{journal}{Phys.
  Rev. B} \textbf{\bibinfo{volume}{83}}, \bibinfo{pages}{132501}
  (\bibinfo{year}{2011}{\natexlab{b}}).

\bibitem[{\citenamefont{Ahilan et~al.}(2008)\citenamefont{Ahilan, Ning, Imai,
  Sefat, Jin, McGuire, Sales, and Mandrus}}]{Imai_08044026}
\bibinfo{author}{\bibfnamefont{K.}~\bibnamefont{Ahilan}},
  \bibinfo{author}{\bibfnamefont{F.~L.} \bibnamefont{Ning}},
  \bibinfo{author}{\bibfnamefont{T.}~\bibnamefont{Imai}},
  \bibinfo{author}{\bibfnamefont{A.~S.} \bibnamefont{Sefat}},
  \bibinfo{author}{\bibfnamefont{R.}~\bibnamefont{Jin}},
  \bibinfo{author}{\bibfnamefont{M.~A.} \bibnamefont{McGuire}},
  \bibinfo{author}{\bibfnamefont{B.~C.} \bibnamefont{Sales}}, \bibnamefont{and}
  \bibinfo{author}{\bibfnamefont{D.}~\bibnamefont{Mandrus}},
  \bibinfo{journal}{Phys. Rev. B} \textbf{\bibinfo{volume}{78}},
  \bibinfo{pages}{100501(R)} (\bibinfo{year}{2008}).

\bibitem[{\citenamefont{Ryan et~al.}(2011)\citenamefont{Ryan,
  Rowan-Weetaluktuk, Cadogan, Hu, Straszheim, Bud'ko, and
  Canfield}}]{Ryan_11030059}
\bibinfo{author}{\bibfnamefont{D.~H.} \bibnamefont{Ryan}},
  \bibinfo{author}{\bibfnamefont{W.~N.} \bibnamefont{Rowan-Weetaluktuk}},
  \bibinfo{author}{\bibfnamefont{J.~M.} \bibnamefont{Cadogan}},
  \bibinfo{author}{\bibfnamefont{R.}~\bibnamefont{Hu}},
  \bibinfo{author}{\bibfnamefont{W.~E.} \bibnamefont{Straszheim}},
  \bibinfo{author}{\bibfnamefont{S.~L.} \bibnamefont{Bud'ko}},
  \bibnamefont{and} \bibinfo{author}{\bibfnamefont{P.~C.}
  \bibnamefont{Canfield}}, \bibinfo{journal}{Phys. Rev. B}
  \textbf{\bibinfo{volume}{83}}, \bibinfo{pages}{104526}
  (\bibinfo{year}{2011}).

\bibitem[{\citenamefont{Zhang et~al.}(2011{\natexlab{b}})\citenamefont{Zhang,
  Xiao, Li, He, Wang, Chen, Normand, Zhang, and Xiang}}]{Zhang_11062706}
\bibinfo{author}{\bibfnamefont{A.~M.} \bibnamefont{Zhang}},
  \bibinfo{author}{\bibfnamefont{J.~H.} \bibnamefont{Xiao}},
  \bibinfo{author}{\bibfnamefont{Y.~S.} \bibnamefont{Li}},
  \bibinfo{author}{\bibfnamefont{J.~B.} \bibnamefont{He}},
  \bibinfo{author}{\bibfnamefont{D.~M.} \bibnamefont{Wang}},
  \bibinfo{author}{\bibfnamefont{G.~F.} \bibnamefont{Chen}},
  \bibinfo{author}{\bibfnamefont{B.}~\bibnamefont{Normand}},
  \bibinfo{author}{\bibfnamefont{Q.~M.} \bibnamefont{Zhang}}, \bibnamefont{and}
  \bibinfo{author}{\bibfnamefont{T.}~\bibnamefont{Xiang}},
  \bibinfo{journal}{arXiv:1106.2706}  (\bibinfo{year}{2011}{\natexlab{b}}).

\bibitem[{\citenamefont{Wang et~al.}(2011{\natexlab{c}})\citenamefont{Wang,
  Fang, D.~X.~Yao, Harriger, Song, Netherton, Zhang, Wang, Stone, Tian
  et~al.}}]{Wang_11054675}
\bibinfo{author}{\bibfnamefont{M.}~\bibnamefont{Wang}},
  \bibinfo{author}{\bibfnamefont{C.}~\bibnamefont{Fang}},
  \bibinfo{author}{\bibfnamefont{G.~T.~T.} \bibnamefont{D.~X.~Yao}},
  \bibinfo{author}{\bibfnamefont{L.~W.} \bibnamefont{Harriger}},
  \bibinfo{author}{\bibfnamefont{Y.}~\bibnamefont{Song}},
  \bibinfo{author}{\bibfnamefont{T.}~\bibnamefont{Netherton}},
  \bibinfo{author}{\bibfnamefont{C.}~\bibnamefont{Zhang}},
  \bibinfo{author}{\bibfnamefont{M.}~\bibnamefont{Wang}},
  \bibinfo{author}{\bibfnamefont{M.~B.} \bibnamefont{Stone}},
  \bibinfo{author}{\bibfnamefont{W.}~\bibnamefont{Tian}}, \bibnamefont{et~al.},
  \bibinfo{journal}{arXiv:1105.4675}  (\bibinfo{year}{2011}{\natexlab{c}}).

\bibitem[{\citenamefont{Zhang et~al.}(2011{\natexlab{c}})\citenamefont{Zhang,
  Lu, and Xiang}}]{Zhang_11024575}
\bibinfo{author}{\bibfnamefont{G.~M.} \bibnamefont{Zhang}},
  \bibinfo{author}{\bibfnamefont{Z.~Y.} \bibnamefont{Lu}}, \bibnamefont{and}
  \bibinfo{author}{\bibfnamefont{T.}~\bibnamefont{Xiang}},
  \bibinfo{journal}{Phys. Rev. B} \textbf{\bibinfo{volume}{84}},
  \bibinfo{pages}{052502} (\bibinfo{year}{2011}{\natexlab{c}}).

\bibitem[{\citenamefont{Normand and Rice}(1997)}]{Normand_PRB}
\bibinfo{author}{\bibfnamefont{B.}~\bibnamefont{Normand}} \bibnamefont{and}
  \bibinfo{author}{\bibfnamefont{T.~M.} \bibnamefont{Rice}},
  \bibinfo{journal}{Phys. Rev. B} \textbf{\bibinfo{volume}{56}},
  \bibinfo{pages}{8760} (\bibinfo{year}{1997}).

\bibitem[{\citenamefont{Zhang et~al.}(2009)\citenamefont{Zhang, Su, Lu, Weng,
  Lee, and Xiang}}]{Zhang_EPL}
\bibinfo{author}{\bibfnamefont{G.}~\bibnamefont{Zhang}},
  \bibinfo{author}{\bibfnamefont{Y.}~\bibnamefont{Su}},
  \bibinfo{author}{\bibfnamefont{Z.}~\bibnamefont{Lu}},
  \bibinfo{author}{\bibfnamefont{Z.}~\bibnamefont{Weng}},
  \bibinfo{author}{\bibfnamefont{D.}~\bibnamefont{Lee}}, \bibnamefont{and}
  \bibinfo{author}{\bibfnamefont{T.}~\bibnamefont{Xiang}},
  \bibinfo{journal}{Europhys. Lett.} \textbf{\bibinfo{volume}{86}},
  \bibinfo{pages}{37006} (\bibinfo{year}{2009}).

\bibitem[{\citenamefont{Wu et~al.}(2008{\natexlab{b}})\citenamefont{Wu,
  Phillips, and CastroNeto}}]{Wu_PRL}
\bibinfo{author}{\bibfnamefont{J.}~\bibnamefont{Wu}},
  \bibinfo{author}{\bibfnamefont{P.}~\bibnamefont{Phillips}}, \bibnamefont{and}
  \bibinfo{author}{\bibfnamefont{A.~H.} \bibnamefont{CastroNeto}},
  \bibinfo{journal}{Phys. Rev. Lett.} \textbf{\bibinfo{volume}{101}},
  \bibinfo{pages}{126401} (\bibinfo{year}{2008}{\natexlab{b}}).

\bibitem[{\citenamefont{de¡¯ Medici et~al.}(2009)\citenamefont{de¡¯ Medici,
  Hassan, Capone, and Dai}}]{Medici_PRL_102}
\bibinfo{author}{\bibfnamefont{L.}~\bibnamefont{de¡¯ Medici}},
  \bibinfo{author}{\bibfnamefont{S.~R.} \bibnamefont{Hassan}},
  \bibinfo{author}{\bibfnamefont{M.}~\bibnamefont{Capone}}, \bibnamefont{and}
  \bibinfo{author}{\bibfnamefont{X.}~\bibnamefont{Dai}},
  \bibinfo{journal}{Phys. Rev. Lett.} \textbf{\bibinfo{volume}{102}},
  \bibinfo{pages}{126401} (\bibinfo{year}{2009}).

\bibitem[{\citenamefont{Arcon et~al.}(2010)}]{Arcon_PRB_82}
\bibinfo{author}{\bibfnamefont{D.}~\bibnamefont{Arcon}} \bibnamefont{et~al.},
  \bibinfo{journal}{Phys. Rev. B} \textbf{\bibinfo{volume}{82}},
  \bibinfo{pages}{140508(R)} (\bibinfo{year}{2010}).

\bibitem[{\citenamefont{Jeglic et~al.}(2010)\citenamefont{Jeglic, Potocnik,
  Klanjsek, Bobnar, Jagodic, Koch, Rosner, Margadonna, Lv, Guloy
  et~al.}}]{Jeglic_LiFeAs_NMR}
\bibinfo{author}{\bibfnamefont{P.}~\bibnamefont{Jeglic}},
  \bibinfo{author}{\bibfnamefont{A.}~\bibnamefont{Potocnik}},
  \bibinfo{author}{\bibfnamefont{M.}~\bibnamefont{Klanjsek}},
  \bibinfo{author}{\bibfnamefont{M.}~\bibnamefont{Bobnar}},
  \bibinfo{author}{\bibfnamefont{M.}~\bibnamefont{Jagodic}},
  \bibinfo{author}{\bibfnamefont{K.}~\bibnamefont{Koch}},
  \bibinfo{author}{\bibfnamefont{H.}~\bibnamefont{Rosner}},
  \bibinfo{author}{\bibfnamefont{S.}~\bibnamefont{Margadonna}},
  \bibinfo{author}{\bibfnamefont{B.}~\bibnamefont{Lv}},
  \bibinfo{author}{\bibfnamefont{A.~M.} \bibnamefont{Guloy}},
  \bibnamefont{et~al.}, \bibinfo{journal}{Phys. Rev. B}
  \textbf{\bibinfo{volume}{81}}, \bibinfo{pages}{140511(R)}
  (\bibinfo{year}{2010}).

\bibitem[{\citenamefont{Li et~al.}(2010)\citenamefont{Li, Ooe, Wang, Liu, Jin,
  Ichioka, and qing Zheng}}]{Li_JPSJ_79}
\bibinfo{author}{\bibfnamefont{Z.}~\bibnamefont{Li}},
  \bibinfo{author}{\bibfnamefont{Y.}~\bibnamefont{Ooe}},
  \bibinfo{author}{\bibfnamefont{X.~C.} \bibnamefont{Wang}},
  \bibinfo{author}{\bibfnamefont{Q.~Q.} \bibnamefont{Liu}},
  \bibinfo{author}{\bibfnamefont{C.~Q.} \bibnamefont{Jin}},
  \bibinfo{author}{\bibfnamefont{M.}~\bibnamefont{Ichioka}}, \bibnamefont{and}
  \bibinfo{author}{\bibfnamefont{G.}~\bibnamefont{qing Zheng}},
  \bibinfo{journal}{J. Phys. Soc. Jpn.} \textbf{\bibinfo{volume}{79}},
  \bibinfo{pages}{083702} (\bibinfo{year}{2010}).

\end{thebibliography}

\end{document}